\documentclass[titlepage,twoside,12pt]{article}
\usepackage{amssymb}
\usepackage{amsfonts}
\begin{document}

{\bf \Large From Reference Frame Relativity to \\ \\ Relativity of Mathematical Models : \\ \\
Relativity Formulas in a Variety of \\ \\ Non-Archimedean Setups} \\

{\bf \large ( part I )} \\ \\

{\bf Elem\'{e}r E Rosinger} \\
Department of Mathematics \\
and Applied Mathematics \\
University of Pretoria \\
Pretoria \\
0002 South Africa \\
eerosinger@hotmail.com \\ \\

{\bf Abstract} \\

Galilean Relativity and Einstein's Special and General Relativity showed that the Laws of
Physics go deeper than their representations in any given reference frame. Thus covariance, or
independence of Laws of Physics with respect to changes of reference frames became a
fundamental principle. So far, all of that has only been expressed within one single
mathematical model, namely, the traditional one built upon the usual continuum of the field
$\mathbb{R}$ of real numbers, since complex numbers, finite dimensional Euclidean spaces, or
infinite dimensional Hilbert spaces, etc., are built upon the real numbers. Here, following
[55], we give one example of how one can go beyond that situation and study what stays the
same and what changes in the Laws of Physics, when one models them within an infinitely large
variety of algebras of scalars constructed rather naturally. Specifically, it is shown that
the Special Relativistic addition of velocities can naturally be considered in any of
infinitely many reduced power algebras, each of them containing the usual field of real
numbers and which, unlike the latter, are non-Archimedean. The nonstandard reals are but one
case of such reduced power algebras, and are as well non-Archimedean. Two surprising and
strange effects of such a study of the Special Relativistic addition of velocities are that
one can easily go beyond the velocity of light, and rather dually, one can as easily end up
frozen in immobility, with zero velocity. Both of these situations, together with many other
ones, are as naturally available, as the usual one within real numbers. \\ \\

{\bf 1. Introduction} \\

There has for longer been an awareness that the exclusive use of the continuum given by the
field $\mathbb{R}$ of real numbers in building up the conventional modelling of Physical
space-time is in fact {\it not} implied by any particular Physical reason, but rather by
Mathematical convenience. Details in this regard can be found in [2,3,5,10,11,21-26,31,48,49,
52,54,55] and the literature cited there. \\
One of the simplest and hardest arguments in this regard has been the observation that only
{\it rational} numbers - thus in $\mathbb{Q}$ and not in the whole of $\mathbb{R}$ - can ever
turn up as results of Physical measurements. And needless to say, the difference between
$\mathbb{Q}$ and $\mathbb{R}$ is considerable, not least since the former is merely countable,
while the latter is not, being in fact uncountable and of the power of the continuum. \\

Needless to say, if we want to replace $\mathbb{R}$, and thus the usual structures built upon
it and used in Physics, with other scalars and corresponding structures, we still need to keep
{\it basic algebraic} operations such as {\it addition, subtraction, multiplication and
division}. However, such operations do {\it not} necessarily imply the use of fields, since
they can be performed as well in the more general algebraic structures called {\it algebras}.
And as well known, and shown in the sequel, there are {\it many more} algebras to use, than
fields. And in fact, there are {\it infinitely many} such algebras which can be constructed in
rather easy ways, thus making their use convenient, see [54,55]. \\

In this way

\begin{quote}

The issue of progressing one step ahead in {\it furthering the Principle of Relativity}, this
time not by mere covariance with respect to reference frames, but by studying the possible
covariance with respect to algebras of scalars in which the Laws of Physics are formulated,
may appear as rather natural.

\end{quote}

A remarkable latest contribution in this regard can be found in [32]. One of the basic
deficiencies of the first von Neumann Hilbert space model of Quantum Mechanics has been the
fact that observables of central importance, such a position, momentum or energy, may lack
eigenstates within the respective Hilbert spaces, [51]. So far, the only rigorous mathematical
approach to this deficiency has been the rigged-Hilbert space formalism. However, in this
approach the respective eigenstates may still fall outside of the Hilbert spaces under
consideration. It is in this regard that [32], by the use of the nonstandard real numbers,
brings a convenient clarification. \\

However, it is important to note the following. The alternative is {\it not at all} restricted
to either using the usual field $\mathbb{R}$ of real numbers, or instead, the nonstandard
field $^*\mathbb{R}$. \\
Indeed, as it happens, there is naturally a far {\it larger} choice at disposal. Namely, one
can study the use of any of infinitely many algebras, among which both the usual reals and the
nonstandard ones are but particular cases. And this infinitely large class of algebras can be
constructed by the {\it reduced power} method presented briefly in the sequel, see also [54,
55]. \\

By the way of the latest contribution in [32], a less well known fact seems to be that, as
early as 1935, von Neumann himself got disappointed in the use of Hilbert spaces in Quantum
Mechanics, as mentioned in one of his letters, [51]. \\

Recent suggestions for going beyond the exclusive use of $\mathbb{R}$ in building models of
Physical space-time have been presented in [2,3,5,10,11,21,23,\\26,48,49,52,54,55], as well as
in the literature mentioned there. \\

Among others, one can envision the replacement of $\mathbb{R}$ by other fields of numbers,
such as for instance, the p-adic fields $\mathbb{Q}_p$, for various prime numbers $p \in
\mathbb{N}$, or even the field $^*\mathbb{R}$ of nonstandard reals. \\
On the other hand, as noted in [54,55], the need to use {\it fields} is not implied by any
specific Physical reason either. Therefore, one could also employ the more general structures
of {\it algebras}, and in this case, as recalled in [55], there is a practically unlimited,
and in fact infinite variety of such algebras which can be constructed easily as {\it reduced
powers} of $\mathbb{R}$. \\

It is important to note that with the conventional use of $\mathbb{R}$, one of the specific
features of $\mathbb{R}$ one has to accept is that, as an ordered field, $\mathbb{R}$ is {\it
Archimedean}. In this regard, two facts can be noted. First, the need to have the Archimedean
property is again not implied by any particular Physical reason. And in fact, this property
often leads to difficulties related to so called "infinities in Physics", difficulties
attempted to be treated by various ad-hoc "re-normalization" procedures. On the other hand,
such difficulties can easily be avoided from the start, if the use of $\mathbb{R}$ is set
aside, since they simply do no longer appear in case one employs instead non-Archimedean
structures, as illustrated in the better known case of nonstandard reals $^*\mathbb{R}$.
Second, the alternatives suggested to $\mathbb{R}$, such as the p-adic fields $\mathbb{Q}_p$
or the infinitely large class of algebras constructed as reduced powers, turn out {\it not} to
be Archimedean. \\

Consequently, lacking any known Physical reason why we should be confined to the use of
Archimedean structures alone, we can equally investigate the way various Laws of Physics may
take shape when non-Archimedean structures are employed. \\

In this regard, in the sequel, we shall study what happens to the law of {\it addition of
velocities} in Special Relativity, when instead of the conventional Archimedean field
$\mathbb{R}$ of real numbers one employs any in the infinite class of algebras of reduced
powers which, as mentioned, are non-Archimedean. \\

One of the unexpected and strange effects of considering the Special Relativistic addition of
velocities in non-Archimedean setup is that one can easily go beyond the velocity of light,
and somewhat dually, one can as easily end up frozen in immobility, with zero velocity, both
of these situations, together with many other ones, being as naturally available, as the usual
one. \\ \\

{\bf 2. Isomorphisms of velocity addition} \\

As shown in [53], and specified briefly in section 4 below, velocity addition in Special
Relativity and Newtonian Mechanics are isomorphic as group operations. \\

Namely, let $c > 0$ be the velocity of light in vacuum. Then, as is well known, in the case of
uniform motion along a straight line, the Special Relativistic addition $*$ of velocities is
given by \\

(SR) $~~~~~~ u * v ~=~ ( u + v ) / ( 1 + u v / c^2 ),~~ u, v \in ( -c, c)$ \\

thus the binary operation $*$~ acts according to \\

$~~~~~~ * : ( -c, c ) \times ( -c, c ) \longrightarrow ( -c, c )$ \\

It follows immediately that \\

(IS1)~~~ $*$~ is associative and commutative \\

(IS2)~~~ $u * v * w ~=~ ( u + v + w + u v w / c^2 ) / ( 1 + ( u v + u w + v w ) / c^2 )$

\hspace*{1.4cm} for $u, v, w \in ( -c, c)$ \\

(IS3)~~~ $u * 0 ~=~ 0 * u ~=~ u,~~ u \in ( -c, c)$ \\

(IS4)~~~ $u * ( -u ) ~=~ ( -u ) * u ~=~ 0,~~ u \in ( -c, c)$ \\

(IS5)~~~ $\partial / \partial u ( u * v ) ~=~ ( 1 - v^2 / c^2 ) / ( 1 + u v / c^2 )^2 > 0,~~
                                                                   u, v \in ( -c, c)$ \\

(IS6)~~~ $\lim_{u,\, v \to c}~ u * v ~=~ c,~~~~ \lim_{u,\, v \to -c}~ u * v ~=~ -c$ \\

Therefore \\

(IS7)~~~ $(~ ( -c, c ), * ~)$ is a commutative group with the neutral

\hspace*{1.4cm} element $0$, while $-u$ is the inverse element of $u \in ( -c, c)$ \\ \\

{\bf 3. Velocity addition in Newtonian Mechanics} \\

As is well known, in the case of uniform motion along a straight line, the addition of
velocities in Newtonian Mechanics is given by \\

(NM) $~~~~~~ x + y,~~ x, y \in \mathbb{R} $ \\

thus it is described by the usual additive group $( \mathbb{R}, + )$ of the real numbers, a
group which is of course commutative, with the neutral element $0$, while $-x$ is the
inverse element of $x \in \mathbb{R}$. \\ \\

{\bf 4. Isomorphisms of the two groups of velocity addition} \\

As shown in [53], the following hold. \\

(IS8)~~~ $(~ ( -c, c ), * ~)$ and $( \mathbb{R}, + )$ are isomorphic groups through the

\hspace*{1.4cm} mappings \\

(IS8.1)~~~ $~~~ \alpha : ( -c, c) \longrightarrow \mathbb{R}$, where \\

$~~~~~~ \alpha ( u ) ~=~ k \ln ( ( c + u ) / ( c - u ) ),~~ u \in ( -c, c)$ \\

and \\

(IS8.2)~~~ $~~~ \beta : \mathbb{R} \longrightarrow ( -c, c)$, where \\

$\beta ( x ) ~=~ c ( e^{ x / k } - 1 ) / ( e^{ x / k } + 1 ),~~ x \in \mathbb{R}$ \\

with \\

(IS8.3) $~~~ k ~=~ c^2 \alpha^\prime ( 0 ) > 0$ \\

(IS9) ~~~ both \,$\alpha$\, and \,$\beta$\, are strictly increasing mappings \\ \\

{\bf 5. Note} \\

The Special Relativistic addition $*$~ of velocities in (SR) is in fact well defined not only
for pairs of velocities \\

$~~~~~~ ( u, v ) \in ( -c, c) \times ( -c, c)$ \\

but also for the {\it larger} set of pairs of velocities \\

$~~~~~~ ( u, v ) \in [ -c, c ] \times [ -c, c ],~~ u v \neq - c^2 $ \\

This corresponds to the fact that in Special Relativity the velocity $c$ of light in vacuum is
supposed to be attainable, namely, by light itself in vacuum. \\

On the other hand, the Newtonian addition $+$~ of velocities (NM) does of course only make
sense Physically for \\

$~~~~~~ ( x, y ) \in \mathbb{R} \times \mathbb{R}$ \\

since infinite velocities are not supposed to be attainable Physically. \\

As for the group isomorphisms $\alpha$ and $\beta$, they only generate mappings between pairs
of velocities in \\

$~~~~~~ ( -c, c) \times ( -c, c) ~\stackrel{\alpha \times \alpha} \longrightarrow~
                                             \mathbb{R} \times \mathbb{R}$ \\

and \\

$~~~~~~ \mathbb{R} \times \mathbb{R}~\stackrel{\beta \times \beta} \longrightarrow~
                                                       ( -c, c) \times ( -c, c)$ \\

thus they do {\it not} cover the cases of addition $u * v$ of special relativistic velocities
$u = -c$ and $v < c$, or $-c < u$, and $v = c$. \\

Consequently, in spite of the group isomorphisms $\alpha$ and $\beta$, there is an {\it
essential} difference between the addition of velocities in Special Relativity, and on the
other hand, Newtonian Mechanics. Indeed, in the latter case, the addition $+$~ is defined on
the {\it open} set $\mathbb{R} \times \mathbb{R}$, while in the former case the addition $*$~
is defined on the set \\

$~~~~~~ \{~ ( u, v ) ~|~ -c \leq u, v \leq c,~~ u v \neq -c^2 ~\}$ \\

which is {\it neither open, nor closed}. \\ \\

{\bf 6. Reduced Power Algebras} \\

We shall now show how the group isomorphisms (IS8.2), (IS8.3) can naturally be extended to
reduced power algebras. \\

First, for convenience, let us recall briefly the general method for constructing an
infinitely large class of algebras obtained as reduced powers, [55]. This {\it reduced power
construction}, in its more general forms, is one of the fundamental tools in Model Theory,
[20,58]. Historically, even if only in a particular case and in an informal manner, it can be
traced back to its use in the 19th century in the classical Cauchy-Bolzano construction of the
field $\mathbb{R}$ of real numbers from the set $\mathbb{Q}$ of rational ones. Various other
familiar instances of the reduced power construction in modern Mathematics can often be
encountered, for instance, when completing metric spaces, or in general, uniform topological
spaces. \\

Let $\Lambda$ be any infinite set, then the power $\mathbb{R}^\Lambda$ is in a natural way an
associative and commutative {\it algebra}. Namely, the elements $\xi \in \mathbb{R}^\Lambda$
can be seen as mappings $\xi : \Lambda \longrightarrow \mathbb{R}$, and as such, they can be
added to, and multiplied with one another point-wise. Namely, if $\xi, \xi\,' : \Lambda
\longrightarrow \mathbb{R}$, then \\

$~~~~~~ ( \xi + \xi\,' ) ( \lambda ) =
                  \xi ( \lambda ) + \xi\,' ( \lambda ),~~~ \lambda \in \Lambda $ \\

$~~~~~~ ( \xi \,.\, \xi\ ) ( \lambda ) =
                  \xi ( \lambda ) \,.\, \xi\,' ( \lambda ),~~~ \lambda \in \Lambda $ \\

In the same way, the elements $\xi \in \mathbb{R}^\Lambda$ can be multiplied with scalars from
$\mathbb{R}$, namely \\

$~~~~~~ ( a \,.\, \xi ) ( \lambda ) = a \,.\, \xi ( \lambda ),~~~
                                              a \in \mathbb{R},~~ \lambda \in \Lambda $ \\

The well known remarkable fact connected with such a power algebra $\mathbb{R}^\Lambda$ is
that there is a one-to-one correspondence between the {\it proper ideals} in it, and on the
other hand, the {\it filters} on the infinite set $\Lambda$, see for instance [54,55] and the
literature cited there. Indeed, this one-to-one correspondence operates as follows \\

(6.1)~~~ $ \begin{array}{l}
                {\cal I} ~~\longmapsto~~ {\cal F}_{\cal I} ~=~
                            \{~ Z(\xi) ~~|~~ \xi \in {\cal I} ~\} \\ \\
                {\cal F} ~~\longmapsto~~ {\cal I}_{\cal F} ~=~
                  \{~ \xi \in \Lambda \longrightarrow \mathbb{R} ~~|~~ Z(\xi) \in {\cal F} ~\}
           \end{array} $ \\

where ${\cal I}$ is an ideal in $\mathbb{R}^\Lambda$, ${\cal F}$ is a filter on $\Lambda$,
while for $ \xi \in \Lambda \longrightarrow \mathbb{R}$, we denoted $Z(\xi) = \{ \lambda \in
\Lambda ~|~ \xi(\lambda) = 0 \}$, that is, the zero set of $\xi$. As is known, the critical
part in (6.1) is that ${\cal F}_{\cal I}$ constitutes a filter on $\Lambda$, see the proof of
(3.7) in [54] for details. \\

The great practical {\it advantage} of the one-to-one correspondence between the proper ideals
in $\mathbb{R}^\Lambda$, and on the other hand, the filters on the infinite set $\Lambda$ is
that the latter are much {\it simpler} mathematical structures. Furthermore, the specific way
reduced powers are constructed, see (6.4), (6.5) below, brings in the power of clarity and
simplicity which made Model Theory such an important branch of modern Mathematics. \\

However, it is important to note that, fortunately, no knowledge of Model Theory is needed in
order to be able to make full use of the reduced power algebras. Indeed, a usual first course
in Algebra, covering such issues as groups, quotient groups, rings and ideals is sufficient.
In this regard, Model Theory comes in only in order to motivate and highlight the naturalness
of the construction which leads to reduced power algebras. \\

Here one further fact should be noted. As is well known, the nonstandard reals $^*\mathbb{R}$
can also be constructed as reduced powers. However, their respective construction is in a way
an extreme case, since it uses the special class of free ultrafilters. An essential resulting
aspect is the so called {\it transfer property}, with the accompanying discrimination between
internal and external entities, which introduces a whole host of technical complications to
deal with, in order to be able to benefit fully from the power of Nonstandard Analysis. \\
In this regard, the use of free ultrafilters is needed in the construction of nonstandard
reals $^*\mathbb{R}$ only in order to obtain them as a {\it totally ordered field}. Otherwise,
if we are ready to use algebras which are not fields, nor totally ordered, we can employ the
much larger class of filters, instead of ultrafilters. Anyhow, even in the case of the
nonstandard reals we end up with the non-Archimedean property. \\

Consequently, the general form of the reduced power construction used here, as well as in [54,
55], does not restrict itself to ultrafilters, and thus avoids the mentioned technical
difficulties related to nonstandard transfer, yet it can benefit from much of the power,
clarity and simplicity familiar in Model Theory. \\

Important properties of the one-to-one correspondence in (6.1) are as follows. Given two
ideals ${\cal I}, {\cal J}$ in $\mathbb{R}^\Lambda$, and two filters ${\cal F}, {\cal G}$ on
$\Lambda$, we have \\

(6.2)~~~ $ \begin{array}{l}
             {\cal I} ~\subseteq~ {\cal J} ~~\Longrightarrow~~
                         {\cal F}_{\cal I} ~\subseteq~ {\cal F}_{\cal J} \\ \\
             {\cal F} ~\subseteq~ {\cal G} ~~\Longrightarrow~~
                         {\cal I}_{\cal F} ~\subseteq~ {\cal I}_{\cal G}
            \end{array} $ \\

Furthermore, the correspondences in (6.1) are idempotent when iterated, namely \\

(6.3)~~~ $ \begin{array}{l}
              {\cal I} ~~\longmapsto~~ {\cal F}_{\cal I}
                      ~~\longmapsto~~ {\cal I}_{{\cal F}_{\cal I}} ~=~ {\cal I} \\ \\
               {\cal F} ~~\longmapsto~~ {\cal I}_{\cal F}
                      ~~\longmapsto~~ {\cal F}_{{\cal I}_{\cal F}} ~=~ {\cal F}
            \end{array} $ \\

It follows that every {\it reduced power algebra} \\

(6.4)~~~ $ A ~=~ \mathbb{R}^\Lambda / {\cal I} $ \\

where ${\cal I}$ is a proper ideal in $\mathbb{R}^\Lambda$, is of the form \\

(6.5)~~~ $ A ~=~ A_{\cal F} ~\stackrel{def}=~ \mathbb{R}^\Lambda / {\cal I}_{\cal F} $ \\

for a suitable unique filter ${\cal F}$ on $\Lambda$. \\

We shall call $\Lambda$ the {\it index set} of the reduced power algebra $A_{\cal F} ~=~
\mathbb{R}^\Lambda / {\cal I}_{\cal F}$, while ${\cal F}$ will be called the {\it generating
filter} which, we recall, is a filter on that index set. \\

Obviously, we can try to relate various reduced power algebra $A_{\cal F} ~=~
\mathbb{R}^\Lambda / {\cal I}_{\cal F}$ to one another, according to the two corresponding
parameters which define them, namely, their infinite index sets $\Lambda$ and their generating
filters ${\cal F}$. \\

We start here by relating them with respect to the latter. Namely, a direct consequence of the
second implication in (6.2) is the following one. Given two filters ${\cal F} \subseteq {\cal
G}$ on $\Lambda$, we have the {\it surjective algebra homomorphism} \\

(6.6)~~~ $ A_{\cal F} \ni \xi + {\cal I}_{\cal F} ~~\longmapsto~~
                                      \xi + {\cal I}_{\cal G} \in A_{\cal G} $ \\

This obviously means that the algebra $A_{\cal G}$ is {\it smaller} than the algebra $A_{\cal
F}$, the precise meaning of it being that \\

(6.6$^*$)~~~ $ A_{\cal G} ~~\mbox{and}~~ A_{\cal F} /
               ( {\cal I}_{\cal G} / {\cal I}_{\cal F} ) ~~\mbox{are isomorphic algebras} $ \\

which follows from the so called {\it third isomorphism theorem for rings}, a classical result
of Algebra. \\

Here we note that in the particular case when the filter ${\cal F}$ on $\Lambda$ is generated
by a nonvoid subset $I \subseteq \Lambda$, that is, when we have \\

(6.7)~~~ ${\cal F} = \{~ J \subseteq \Lambda ~|~ J \supseteq I ~\} $ \\

then it follows easily that \\

(6.8)~~~ $ A_{\cal F} ~=~ \mathbb{R}^I $ \\

which means that we do {\it not} in fact have a reduced power algebra, but only a power
algebra. \\

For instance, in case $I$ is finite and has $n \geq 1$ elements, then $A_{\cal F} =
\mathbb{R}^n$ is in fact the usual n-dimensional Euclidean space. \\

Consequently, in order to avoid such a degenerate case of reduced power algebras, we have to
avoid the filters of the form (6.7). This can be done easily, since such filters are obviously
characterized by the property \\

(6.9)~~~ $ \bigcap_{\,J\, \in\, {\cal F}}~ J ~=~ I \neq \phi $ \\

It follows that we shall only be interested in filters  ${\cal F}$ on $\Lambda$ which satisfy
the condition \\

(6.10)~~~ $ \bigcap_{\,J\, \in\, {\cal F}}~ J ~=~ \phi $ \\

or equivalently \\

(6.11)~~~ $ \begin{array}{l}
                \forall~~ \lambda \in \Lambda ~: \\ \\
                \exists~~ J_\lambda \in {\cal F} ~: \\ \\
                ~~~ \lambda \notin J_\lambda
            \end{array} $ \\

which is further equivalent with \\

(6.12)~~~ $ \begin{array}{l}
                \forall~~ I \subset \Lambda,~~ I ~~\mbox{finite} ~: \\ \\
                ~~~ \Lambda \setminus I \in {\cal F}
            \end{array} $ \\

We recall now that the {\it Frech\'{e}t filter} on $\Lambda$ is given by \\

(6.13)~~~ $ {\cal F}re ( \Lambda ) ~=~ \{~ \Lambda \setminus I ~~|~~
                             I \subset \Lambda,~~ I ~~\mbox{finite} ~\} $ \\

In this way, condition (6.10) - which we shall ask from now on about all filters ${\cal F}$ on
$\Lambda$ - can be written equivalently as \\

(6.14)~~~ $ {\cal F}re ( \Lambda ) ~\subseteq~ {\cal F} $ \\

This in particular means that \\

(6.14$^*$)~~~ $ \begin{array}{l}
                    \forall~~ I \in {\cal F} ~: \\ \\
                    ~~~ I ~~\mbox{is infinite}
                \end{array} $ \\

Indeed, if we have a finite $I \in {\cal F}$, then $\Lambda \setminus I \in {\cal F}re (
\Lambda )$, hence (6.14) gives $\Lambda \setminus I \in {\cal F}$. But $I \cap ( \Lambda
\setminus I ) = \phi$, and one of the axioms of filters is contradicted. \\

In view of (6.6), it follows that all reduced power algebras considered from now on will be
{\it homomorphic images} of the reduced power algebra $A_{{\cal F}re ( \Lambda )}$, through
the surjective algebra homomorphisms \\

(6.15)~~~ $ A_{{\cal F}re ( \Lambda )} \ni \xi + {\cal I}_{{\cal F}re ( \Lambda )}
                             ~~\longmapsto~~ \xi + {\cal I}_{\cal F} \in A_{\cal F} $ \\

or in view of (6.6$^*$), we have the {\it isomorphic algebras} \\

(6.15$^*$)~~~ $ A_{\cal F},~~~ A_{{\cal F}re ( \Lambda )} /
                        ( {\cal I}_{\cal F} /{\cal I}_{{\cal F}re ( \Lambda )} ) $ \\

Let us note that the {\it nonstandard reals}~ $^*\mathbb{R}$ are a particular case of the
above reduced power algebras (6.4). Indeed, $^*\mathbb{R}$ can be defined by using {\it free
ultrafilters} ${\cal F}$ on $\Lambda$, that is, ultrafilters which satisfy (6.10), or
equivalently (6.14). \\

We note that the field of real numbers $\mathbb{R}$ can be embedded naturally in each of the
reduced power algebras (6.4), by the {\it injective algebra homomorphism} \\

(6.16)~~~ $ \mathbb{R} \ni x ~~\longmapsto~~ \xi_x + {\cal I} \in A $ \\

where $\xi_x ( \lambda ) = x$, for $\lambda \in \Lambda$. Indeed, if $\xi_x \in {\cal I}$
and $\xi \neq 0$, then the ideal ${\cal I}$ must contain $x_1$, which means that it is {\it
not} a proper ideal, thus contradicting the assumption on it. \\

For simplicity of notation, we may write $\xi_x = x$, for $x \in \mathbb{R}$, thus (6.16) can
take the form \\

(6.17)~~~ $ \mathbb{R} \ni x ~~\longmapsto~~ x = \xi_x + {\cal I} \in A $ \\

which in view of the injectivity of this mapping, we may further simplify to \\

(6.18)~~~ $ \mathbb{R} \ni x ~~\longmapsto~~ x \in A $ \\

in other words, to the {\it algebra embedding} \\

(6.18$^*$)~~~ $ \mathbb{R} ~\subsetneqq~ A $ \\

There is also the issue to relate reduced power algebras corresponding to different {\it index
sets}. Namely, let $\Lambda \subseteq \Gamma$ be two infinite index sets. Then we have the
obvious {\it surjective algebra homomorphism} \\

(6.19)~~~ $ \mathbb{R}^\Gamma \ni \xi ~~\longmapsto~~
                       \xi|_{\,\Lambda} \in \mathbb{R}^\Lambda $ \\

since the elements $\xi \in \mathbb{R}^\Gamma$ can be seen as mappings $\xi : \Gamma
\longrightarrow \mathbb{R}$. Consequently, given any ideal ${\cal I}$ in $\mathbb{R}^\Gamma$,
we can associate with it the ideal in $\mathbb{R}^\Lambda$, given by \\

(6.20)~~~ $ {\cal I}|_{\,\Lambda} ~=~ \{~ \xi|_{\,\Lambda} ~~|~~ \xi \in {\cal I} ~\} $ \\

As it happens, however, such an ideal ${\cal I}|_{\,\Lambda}$ need not always be a proper
ideal in $\mathbb{R}^\Lambda$, even if ${\cal I}$ is a proper ideal in $\mathbb{R}^\Gamma$.
For instance, if we take $\gamma \in \Gamma \setminus \Lambda$, and consider the proper ideal
in $\mathbb{R}^\Gamma$ given by ${\cal I} = \{ \xi \in \mathbb{R}^\Gamma ~|~ \xi(\gamma) = 0 ~\}$,
then we obtain ${\cal I}|_{\,\Lambda} = \mathbb{R}^\Lambda$, which is not a proper ideal in
$\mathbb{R}^\Lambda$. \\

We can avoid that difficulty by noting the following. Given a filter ${\cal F}$ on $\Gamma$
which satisfies (6.14), that is, ${\cal F}re ( \Gamma ) \subseteq~ {\cal F}$, then \\

(6.21)~~~ $ {\cal F}|_{\,\Lambda} ~=~ \{~ I \cap \Lambda ~~|~~ I \in {\cal F} ~\} $ \\

satisfies the corresponding version of (6.14), namely ${\cal F}re ( \Lambda ) ~\subseteq~
{\cal F}|_{\,\Lambda}$. Indeed, let us take $J \subseteq \Lambda$ such that $\Lambda \setminus
J$ is finite. Then clearly $\Gamma \setminus ( J \cup ( \Gamma \setminus \Lambda ) )$ is
finite, hence $J \cup ( \Gamma \setminus \Lambda ) \in {\cal F}$. However, $J = ( J \cup (
\Gamma \setminus \Lambda ) ) \cap \Lambda )$, thus $J \in {\cal F}|_{\,\Lambda}$. \\
Now in order for ${\cal F}|_{\,\Lambda}$ to be a filter on $\Lambda$, it suffices to show that
$\phi \notin {\cal F}|_{\,\Lambda}$. Assume on the contrary that for some $I \in {\cal F}$ we
have $I \cap \Lambda = \phi$, then $I \subseteq \Gamma \setminus \Lambda$, thus $\Lambda
\notin {\cal F}$. \\

It follows that \\

(6.22)~~~ $ {\cal F}|_{\,\Lambda} ~~\mbox{is a filter on}~ \Lambda
                         ~\mbox{which satisfies}~ (6.14)
                                    ~~\Longleftrightarrow~~ \Lambda \in {\cal F} $ \\

In view of (6.19) - (6.22), for every filter ${\cal F}$ on $\Gamma$, such that \\

(6.23)~~~ $ \Lambda \in {\cal F} $ \\

we obtain the {\it surjective algebra homomorphism} \\

(6.24)~~~ $ A_{\cal F} ~=~ \mathbb{R}^\Gamma / {\cal I}_{\cal F} \ni
               \xi + {\cal I}_{\cal F} ~~\longmapsto~~
                 \xi|_{\,\Lambda} + {\cal I}_{{\cal F}|_{\,\Lambda}} \in
                    A_{{\cal F}|_{\,\Lambda}} ~=~
                           \mathbb{R}^\Lambda / {\cal I}_{{\cal F}|_{\,\Lambda}} $ \\

and in particular, we have the following relation between the respective proper ideals \\

(6.25)~~~ $ ( {\cal I}_{\cal F} )|_{\,\Lambda} ~=~ {\cal I}_{{\cal F}|_{\,\Lambda}} $ \\ \\

{\bf 7. Zero Divisors and the Archimedean Property} \\

It is an elementary fact of Algebra that a quotient algebra (6.4) has {\it zero divisors},
unless the ideal ${\cal I}$ is {\it prime}. A particular case of that is when a quotient
algebra (6.4) is a {\it field}, which is characterized by the ideal ${\cal I}$ being {\it
maximal}. And in view of (6.5), (6.2), this means that the filter ${\cal F}$ generating such
an ideal must be an {\it ultrafilter}, see for details [54,55]. \\

On the other hand, {\it none} of the reduced power algebras (6.5) which correspond to filters
satisfying (6.14) are {\it Archimedean}. And that includes the nonstandard reals
$^*\mathbb{R}$ as well. \\

In this regard, let us first we note that on reduced power algebras (6.5), one can naturally
define a {\it partial order} as follows. Given two elements $\xi + {\cal I}_{\cal F},~ \eta +
{\cal I}_{\cal F} \in A_{\cal F} = \mathbb{R}^\Lambda / {\cal I}_{\cal F}$, we define \\

(7.1)~~~ $ \xi + {\cal I}_{\cal F} ~\leq~ \eta + {\cal I}_{\cal F} ~~\Longleftrightarrow~~
                    \{~ \lambda \in \Lambda ~~|~~
                           \xi ( \lambda ) ~\leq~ \eta ( \lambda ) ~\} \in {\cal F} $ \\

Now, with this partial order, the algebra $A_{\cal F}$ would be {\it Archimedean}, if and only
if \\

(7.2)~~~ $ \begin{array}{l}
               \exists~~ \upsilon + {\cal I}_{\cal F} \in A_{\cal F},~
                                            \upsilon + {\cal I}_{\cal F} \geq 0 ~: \\ \\
               \forall~~ \xi + {\cal I}_{\cal F} \in A_{\cal F},~
                                            \xi + {\cal I}_{\cal F} \geq 0 ~: \\ \\
               \exists~~ n \in \mathbb{N} ~: \\ \\
               ~~~  \xi + {\cal I}_{\cal F} ~\leq~ n \upsilon + {\cal I}_{\cal F}
            \end{array} $ \\ \\

However, in view of (6.14), we can take an infinite $I \in {\cal F}$. Thus we can define a
mapping $\omega : \Lambda \longrightarrow \mathbb{R}$ which is unbounded from above on $I$.
And in this case taking $\xi + {\cal I}_{\cal F} = ( \upsilon + \omega ) + {\cal I}_{\cal F}$,
it follows easily that condition (7.2) is not satisfied. \\

We note that the reduced power algebras $A_{\cal F}$ in (6.5) are Archimedean only in the
degenerate case (6.7), (6.8), when in addition the respective sets $I$ are {\it finite}, thus
as noted, the respective algebras reduce to finite dimensional Euclidean spaces. \\

Further we note that the partial order (7.1) on the algebras $A_{\cal F}$ is in general {\it
not} a total order. In this regard, we have \\

{\bf Proposition 1.} \\

Let ${\cal F}$ be a filter on $\Lambda$ satisfying (6.14). Then the partial order (7.1) is a
total order on the reduced power algebra $A_{\cal F}$, if and only if ${\cal F}$ is an
ultrafilter. In that case we have $A_{\cal F} = \,^*\mathbb{R}$, that is, the reduced power
algebra $A_{\cal F}$ is the field $^*\mathbb{R}$ of nonstandard reals. \\

{\bf Proof.} \\

Let us take any partition $\Lambda = \Lambda_0 \cup \Lambda_1$ into two infinite subsets and
take $\xi_0,~ \xi_1 : \Lambda \longrightarrow \mathbb{R}$ as the characteristic functions of
$\Lambda_0$ and $\Lambda_1$, respectively. Then obviously \\

(7.3)~~~ $ \{ \lambda \in \Lambda ~|~ \xi_0 ( \lambda ) < \xi_1 ( \lambda ) \} = \Lambda_1,~~~
         \{ \lambda \in \Lambda ~|~ \xi_0 ( \lambda ) > \xi_1 ( \lambda ) \} = \Lambda_0 $ \\

and in general $\Lambda_0,~ \Lambda_1 \notin {\cal F}$, like for instance, when ${\cal F} =
{\cal F}re ( \Lambda )$. Thus in view of (7.1), in general, we cannot have in $A_{\cal F}$
either of the inequalities \\

(7.4)~~~ $ \xi_0 + {\cal I}_{\cal F} ~\leq~ \xi_1 + {\cal I}_{\cal F},~~~
            \xi_0 + {\cal I}_{\cal F} ~\geq~ \xi_1 + {\cal I}_{\cal F} $ \\

Clearly, no filter on $\Lambda$ can simultaneously contain both $\Lambda_0$ and $\Lambda_1$,
thus both of the above inequalities (7.4) can never hold simultaneously. \\

However, in case ${\cal F}$ is an ultrafilter satisfying (6.14), thus we are in the particular
situation when $A_{\cal F} = \,^*\mathbb{R}$, that is, the reduced power algebra $A_{\cal F}$
is the field $^*\mathbb{R}$ of nonstandard reals, then according to a property of ultrafilters,
we must have either $\Lambda_0 \in {\cal F}$, or $\Lambda_1 \in {\cal F}$. Therefore, one and
only one of the above two inequalities in (7.4) holds. \\

Conversely, let (7.1) be a total order on $A_{\cal F}$. Then one and only one of the
inequalities (7.4) must hold. Let us assume that it is the case of the first one of them. Then
(7.1), (7.3) imply that $\Lambda_1 \in {\cal F}$. Obviously, in the other case we obtain that
$\Lambda_0 \in {\cal F}$. \\

In this way, whenever we are given a partition $\Lambda = \Lambda_0 \cup \Lambda_1$ into two
infinite subsets, one of them must belong to the filter ${\cal F}$. \\
If on the other hand, in the partition $\Lambda = \Lambda_0 \cup \Lambda_1$ one of the sets is
finite, then in view of (6.14) the other must belong to the filter ${\cal F}$. And since
$\Lambda$ is supposed to be infinite, both sets in the partition cannot be finite. \\

Thus we can conclude that the filter ${\cal F}$ is indeed an ultrafilter on $\Lambda$, since
its above property related to partition characterizes ultrafilters. \\ \\

\end{document}